Helping students become proficient problem solvers Part II: An example from waves

Chandralekha Singh[1] and Alexandru Maries[2]
[1]Department of Physics and Astronomy, University of Pittsburgh, Pittsburgh, PA, 15260, USA
[2] Department of Physics, University of Cincinnati, Cincinnati, OH 45221, USA

Helping students become proficient problem-solvers is one of the primary goals of physics courses. In part 1 of this article, we summarized the vast research on problem-solving relevant for physics instruction, and here we discuss a concrete example of problem solving in the context of waves from introductory physics. The goal of this research was to investigate how drawing of diagrams affects students' problem-solving performance. An introductory class was broken up into three recitations which received different instructions related to diagrams on their weekly quizzes: one group was provided a diagram, another was asked to draw one, and the third was the comparison group which was given no instructions about diagrams. We find that students who were provided a diagram performed significantly worse than students in the other two groups. Furthermore, we find that irrespective of the condition, students who drew diagrams as part of the problem-solving process performed better overall despite primarily using a mathematical approach to solving the problem. Lastly, we conducted think-aloud interviews with students who solved the same problem to further understand their solution approaches as well as how drawing a diagram is useful even in situations where a primarily mathematical approach is used.

## I. Introduction

Introductory physics is a challenging subject to learn at least partly because students rarely associate the abstract concepts they study in physics with more concrete representations that facilitate understanding without an explicit instructional strategy aimed to aid them. With guidance and scaffolding support, students often have formula-centric problem-solving strategies that may not require understanding of physical concepts. Unfortunately, these inferior strategies can be rewarded in traditionally taught introductory physics courses [1].

The benefits of multiple representations in problem solving have long been documented in physics education research [1-20], and therefore, many researchers have developed instructional strategies that place explicit emphasis on multiple representations [3, 4, 6, 20, 21] while other researchers developed strategies with implicit focus on multiple representations [22-27]. Van Heuvelen's approach, for example, [3] starts by ensuring that students explore the qualitative nature of concepts by using a variety of representations of a concept in a familiar setting before adding the complexities of mathematics. Many other researchers have emphasized the importance of students becoming facile in translating between different representations of concepts [6, 28] and that significant positive learning occurs when students develop facility in the use of multiple representation [29]. However, careful attention must be paid to instructional use of diverse representational modes as specific learning difficulties may arise as a consequence [30] because students can approach the same problem posed in a different representation differently without support [30, 31].

One representation useful in the initial conceptual analysis and planning stage of a solution is a schematic diagram of the physical situation presented in the problem. Diagrammatic representations have been shown to be superior to exclusively employing verbal representations when solving problems [6]. It is therefore not surprising that physics experts automatically employ diagrams in attempting to solve problems [3, 4, 22, 32, 33]. However, introductory physics students need explicit help understanding that drawing a diagram is an important step in organizing and simplifying the given information into a representation which is more suitable to further analysis [34]. Therefore, many researchers who have developed strategies for helping students learn effective problem-solving skills attempt to make students realize how important the step of drawing a diagram is in solving a physics problem. In Newtonian mechanics, Reif [1, 2] has suggested that several diagrams be drawn: one diagram of the problem description which includes all objects and one diagram for each subsystem that needs to be considered separately. Also, he detailed concrete steps that students need to take in order to draw those diagrams: (a) describe both motions and interactions, (b) identify interacting objects before forces, (c) separate long range and contact interactions and (d) label contact points by the magnitude of the action-reaction pair of forces. Other researchers who have emphasized, among other things, the importance of diagrams in their approach to helping students learn problem solving skills have found significant improvements in students' problem-solving methods [3]. Previous research shows that students who draw diagrams even if they are not rewarded for them are more successful problem solvers [35]. An investigation into how spontaneous drawing of free body diagrams (FBDs) [36] affects problem solving [37] shows that only drawing correct FBDs improves a student's score and that students who draw incorrect FBDs do not perform better than students who draw no diagrams.

The study presented here relates to student understanding of the concept of mechanical waves as it relates to harmonics of standing waves in cylindrical tubes. Conceptions of mechanical waves have been researched in young children [38, 39], middle school and high-school students [40-43], introductory undergraduate students [44, 45], advanced undergraduate students [46], and physics teacher candidates [47]. Eshach and Schwartz [40] investigated whether Reiner's [48] earlier finding that the initial knowledge that students bring to the study of science is often "substance based" (which Reiner termed "substance schema") also holds for mechanical waves. They found that students do hold this view in some respects. However, sometimes students perceive the "substance" that they associate with sound waves differently from other "regular" substances. Later Eshach and colleagues [49] developed a statistical tool for estimating students' mental models. Wittmann [50] also found that students often use reasoning that is focused on object-like properties when discussing waves which can be problematic to the goal of shifting student understanding of mechanical waves from a substance-based ontology to a sequence of events. In addition, Wittmann and Hrepic [51, 52] were interested in identifying students' mental models of mechanical waves and how knowledge of these mental models can be used to improve students' understanding of mechanical waves. One tool for identifying these mental models is Wittmann's Wave Diagnostic Test [53], an open-ended questionnaire. Tonghchai et al., on the other hand, developed a multiple-choice assessment tool for mechanical waves [54] and used it [55] to evaluate the consistency of students' conceptions. They found that students solve problems involving mechanical waves across different contexts inconsistently, much like what other researchers have found for other areas of physics [56-60]. Student understanding of other types of waves such as light waves [61] and electromagnetic waves [62] has also been researched. However, the use of multiple representations and its role to understanding mechanical waves has not been researched in much depth. Among the few who have investigated the role of non-verbal representations to understanding sound waves, Eshach and Schwartz [40] found that students have a variety of non-verbal representations that they employ while explaining their understanding of sound waves. During the interviews they conducted with high-school students, they allowed them to draw or gesticulate to explicate their reasoning. They concluded this research by suggesting that these non-verbal representations students use could be a good starting point to help them construct the correct visual representations needed to fully understand wave propagation phenomena.

Here, we describe an investigation about how algebra-based introductory physics students' performance on a problem (given in a quiz) related to standing waves in a tube is affected when students are given a diagram as opposed to when they are asked to draw a diagram (without being more specific than that). The performance of these students was also contrasted with that of a comparison group which was not given any instructions related to diagrams when solving the same problem related to standing waves. Moreover, a second similar problem was given in a midterm exam for which all introductory physics students received no instructions regarding diagrams. The primary goals of this study were to investigate how the different interventions impact students' problem-solving performance as well as to investigate more broadly how drawing a diagram impacts students' problem-solving performance, particularly in situations where their chosen solution primarily involved manipulations of equations. We found that 1) students who were explicitly asked to draw diagrams drew more productive diagrams than students in the other two groups and 2) students who were provided a diagram performed significantly worse than students in the other two groups. Moreover, both in the quiz problem and in the midterm problem, students who used a mathematical approach, but also drew productive diagrams, performed better than students who used a mathematical approach without drawing a productive diagram. In addition, we found that many students employing the mathematical approach had difficulties manipulating two equations symbolically in the context of solving the problem involving standing waves in a tube. In order to better understand these findings, we conducted think-aloud interviews with eight students enrolled in another algebra-based introductory physics course. The interviews were helpful in furnishing or corroborating possible interpretations of the quantitative results from in-class quizzes.

## II.   Methodology

### a.  Participants

A class of 118 introductory physics students in an algebra-based course was broken up into three different recitations. All recitations were taught in a traditional manner in which the teaching assistant (TA) worked out problems similar to the homework problems and then gave a 15-20 minute quiz at the end of the recitation. Students in all recitations attended the same lectures, were assigned the same homework, and had the same exams and quizzes. The students were majoring in biological sciences or health-related majors and were primarily juniors and seniors. Also, this was the second semester of their introductory physics sequence.

In addition to the introductory students, one of the problems used in this study was also given to 26 first year physics graduate students (physics experts for this study) enrolled in a TA training course in order to assess how often physics experts use the diagrammatic approach to solving the problem (this approach was hypothesized to be a more expert-like approach). We also were interested in comparing the average score that graduate students obtained on this problem with that of introductory students. In order to make sure that the graduate students did not use a diagram simply because they did not remember the relevant equations, those equations were provided.

### b.  Interventions used this study

In the recitation quizzes throughout the semester, the three groups were given the same problems but with the following interventions: in each quiz problem, the first intervention group, which we refer to as the "prompt only group" or "PO", was given an explicit prompt or instruction to draw a diagram along with the problem statement; the second intervention group (referred to as the "diagram only group" or "DO") was given a diagram drawn by the instructor that was meant to aid in solving the problem and the third group was the comparison group which was not given any diagram or explicit instruction to draw a diagram with the problem statement ("no support group" or "NS").

The sizes of the different recitation groups varied from 22 to 55 students because the students were not assigned a particular recitation so they could go to whichever recitation they wanted. For the same reason, the size of each recitation group also varied from week to week, although not as drastically because most students (≈ 80%) would stick with a particular recitation. Furthermore, each intervention was not matched to a particular recitation. For example, in one week, NS was the Tuesday recitation while another week, NS was a different recitation section. This is important because it implies that individual students were subjected to different interventions from week to week and we do not expect cumulative effects due to the same group of students always being subjected to the same intervention.

c. *Problem used, correct solution and grading criteria*

Both problems used in this study involve standing waves in tubes. One of the problems was given in a quiz in which the interventions were implemented and the other was given in a midterm exam in which there was no intervention and all students received the same instructions.

- Quiz problem: "A tube with air is open at only one end and has a length of 1.5 m. This tube sustains a standing wave at its third harmonic. What is the distance between a node and the adjacent antinode?". We note that students in the DO group were provided with a diagram of an empty tube in addition to this problem statement. Students in the PO intervention were explicitly asked to draw a diagram after the above problem statement, and students in the comparison group (NS) only saw the problem statement.
- Midterm problem: The midterm exam problem was identical to the quiz problem except that the tube was open at both ends instead of just one end. All students received the problem statement only. No diagram or prompt to draw a diagram was included.

There are two approaches to solving the quiz problem (the midterm problem can also be solved by employing a very similar strategy for a tube that is open at both ends). One strategy is to draw the standing wave in question as shown in Fig. 1. (We note that in Fig. 1 we have chosen to represent the wave by displacement rather than density or pressure variations of the gas since displacement was emphasized in the class.) Then, for example, one can identify that three node to antinode distances fit in the tube, and since the length of the tube is $L = 1.5$ m, the distance between a node and the adjacent antinode is $1.5/3 = 0.5$ m. This diagrammatic approach is the more expert-like approach because it requires understanding of a physics concept in its diagrammatic representation (third harmonic of a standing wave) and how it applies to a tube which is open at only one end (node at the closed end and antinode at the open end).

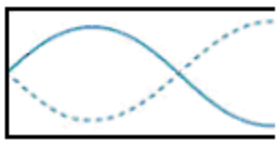

**Figure 1:** 3rd harmonic of a standing wave in a tube open at only one end.

The second approach to solving this problem (very similar to the second approach for the midterm problem) is to use the equation for the frequency of the $n^{th}$ harmonic of a standing wave in a tube of length $L$ open at only one end, $f_n = \frac{nv}{4L}$, and the relation between the speed $v$, frequency $f$ and wavelength $\lambda$ of a wave, namely $v = f\lambda$. Then, one can solve for the wavelength given $L$ and $n$ and finally divide the wavelength by 4 to obtain the distance between a node and the adjacent antinode. We refer to this latter approach as the "mathematical" approach because it does not necessarily require understanding the physics principles involved: students can manipulate the two equations and solve for the target variable without much understanding of the underlying physics concepts.

Students in DO were not provided the diagram in Fig. 1 because it would have greatly reduced the difficulty of the problem. Instead, they were given a partial diagram: the cross-section of the tube without a representation of the wave in the tube. It was intended that students would regard the partial diagram as a hint to complete it and be more likely to follow the expert like diagrammatic approach.

In order to ensure homogeneity of grading, rubrics were developed for each problem analyzed and the rubrics were used to ensure that there was at least 90% inter-rater-reliability between two different raters who independently graded 20% of the data (the rest of the data was graded by one rater). The development of the rubric for each problem went through an iterative process. During the development of the rubric, the two raters discussed students' scores separately from the ones obtained using the prior version of the rubric and adjusted the rubric if it was agreed that the version of the rubric was too stringent or too generous. After each adjustment to the rubric, all the students' scores were computed again using the improved rubric. Since there are two approaches to the solution of these problems, one primarily diagrammatic and another primarily based on mathematical manipulations, rubrics were developed to score the performance of students employing each approach. The summary of the rubric used to score students out of 10 points who chose the mathematical approach (both in the quiz and midterm problem) is shown in Table 1.

It is important to emphasize that a tube open only at one end can only sustain the odd harmonics, therefore $n$ in the formula $f_n = \frac{nv}{4L}$ takes only odd values. However, the third harmonic corresponds to $n = 3$ (and not the third possible value for $n$, namely 5) due to the convention that the frequency of the $n^{th}$ harmonic of a standing wave must be $n$ times the fundamental frequency. This is a common source of confusion even among graduate students. We surveyed several physics graduate students and found that most of them associated the third harmonic with the third possible value for $n$, namely 5, which is incorrect based on the convention. Similarly, they incorrectly thought that the diagrammatic representation of the third harmonic corresponds to the third possible way of drawing a standing wave in a tube open at only one end, which instead corresponds to the fifth harmonic due to the same convention. Since many experts confuse the fifth and third harmonics, the researchers considered that one should not penalize introductory students for this confusion. However, we also performed data analysis in an alternative manner in which one point (out of a maximum of 10) was taken off for mistaking the third harmonic with the fifth or using $n = 5$. All of the results are qualitatively identical: every comparison which yielded a statistically significant difference in one instance (not taking off points for this mistake) also yielded a statistically significant difference in the other instance (taking off one point for this mistake). Here, we report the results obtained by using the grading approach which does not penalize students for mistaking the third harmonic with the fifth.

Table 1 shows that there are two parts to the rubric: Correct and Incorrect Ideas. Table 1 also shows that in the Correct Ideas part, the problem was divided into different sections and points were assigned to each section (10 maximum points). Each student starts out with 10 points and in the Incorrect Ideas part, the common mistakes students made in each section and the number of points that were deducted for each of those mistakes are listed. It is important to note that each mistake is connected to a particular section (the mistakes labeled 1 and 2 are for the first and second sections, respectively, the two mistakes labeled 3.1 and 3.2 are for the third section and so on) and that for each section, the rubric cannot be used to subtract more points than that section is worth. For example, the two mistakes in section 3 (3.1 and 3.2) are mutually exclusive. Similarly, mistake 4.1 is exclusive with all other mistakes in section 4 and mistakes 4.3 and 4.4 are mutually exclusive. Finally, if the mistake a student made was not common and not in the rubric, it would correspond to the mistake labeled as 4.4.

A rubric was also developed to score the performance of students employing the diagrammatic approach. The summary of the rubric is shown in Table 2.

**TABLE 1:** Summary of the rubric used to score the performance in the quiz of students employing the mathematical approach out of 10 points.

| | **Correct Ideas** | |
|---|---|---|
| Section 1 | Used equation provided $f_n = \frac{nv}{4L}$ | 1 p |
| Section 2 | Chose $n = 3$ or $n = 5$ | 1 p |
| Section 3 | Wrote $v = f\lambda$ | 3 p |
| Section 4 | Solved for $\lambda$ correctly | 2 p |
| Section 5 | Calculated distance between node and antinode by dividing $\lambda$ by 4 | 2 p |
| Section 6 | Correct unit for answer | 1 p |
| | **Incorrect Ideas** | |
| Section 1 | Used incorrect equation | -1 p |
| Section 2 | Chose value for *n* other than 3 or 5 | -1 p |
| Section 3 | 3.1 Did not write $v = f\lambda$ | -3 p |
| | 3.2 Tried to write down $v = f\lambda$, but made a mistake (i.e., wrote something like $v = f/\lambda$) | -2 p |
| Section 4 | 4.1 Did not solve for $\lambda$ | -2 p |
| | 4.2 Used a value for $v$ other than that for sound wave | -1 p |
| | 4.3 Made an error and obtained incorrect $\lambda$ | -1 p |
| | 4.4 Unclear how $\lambda$ was calculated or other error | -1 p |
| Section 5 | Did not divide $\lambda$ by 4 to obtain the answer or did not obtain an answer | -2 p |
| Section 6 | Incorrect units | -1 p |

**TABLE 2:** Summary of the rubric used to score the performance in the quiz of students employing the diagrammatic approach, out of 10 points.

| | **Correct Ideas** | |
|---|---|---|
| Section 1 | Drew a diagram of the correct standing wave | 4 p |
| Section 2 | Used diagram correctly to obtain the answer | 5 p |
| Section 3 | Correct units for answer | 1 p |
| | **Incorrect Ideas** | |
| Section 1 | 1.1 Diagram is a sinusoidal wave that does not clearly indicate locations of nodes and antinodes | -1 p |
| | 1.2 Diagram has either two nodes or two antinodes at the endpoints | -2 p |
| | 1.3 Diagram does not represent the third or fifth harmonic* (if endpoints are a node and an antinode) | -1 p |
| | 1.4 Diagram does not represent the third harmonic** (if endpoints are both nodes or both antinodes) | -1 p |
| Section 2 | 2.1 Answer found is not the distance between a node and an antinode, nor the distance between two nodes (based on student's diagram) | -4 p |
| | 2.2 Used diagram correctly, but found the distance between two nodes | -2 p |
| | 2.3 Unclear how answer was obtained or other error | -1 p |
| Section 3 | Incorrect units | -1 p |

\* Due to the confusion even amongst many experts between the third and fifth harmonic for a standing wave in a tube open at only one end, both diagrams were considered correct.

\*\* For the case when the tube is open or closed at both ends, experts do not have difficulties because all harmonics are possible, including even harmonics, therefore, it was considered that students should know in those instances how the third harmonic should be drawn.

The basic form of the summary of the rubric shown in Table 2 is the same as the one shown in Table 1; it has the same two main parts (Correct and Incorrect Ideas), the problem is broken up into sections and the common mistakes students made in each section are listed. In section 1, mistakes 1.4 and 1.3 are mutually exclusive and in section 2, mistakes 2.1 and 2.2 are mutually exclusive. Similar to the rubric used for the mathematical approach, we left ourselves a small window (labeled 2.3) if a mistake of a student was not explicitly in the rubric (a very rare occurrence, less than 5% of the cases).

We note that the rubrics are designed to be similar in terms of penalizing for mistakes that could be considered as analogous. For example, the rubric used for the mathematical approach treats the cases $n = 3$ and $n = 5$ as both correct because of the confusion of experts. Similarly, the rubric used for the diagrammatic approach does not penalize of student for drawing the fifth harmonic instead of the third of a standing wave in a tube open at only one end. Since this confusion was not penalized in one rubric, it was also not penalized in the other. Another example is provided by the analogy between the mistake of section 5 in the mathematical rubric and the mistake labeled 2.2 in the diagrammatic one. Students who use the mathematical approach and find the wavelength correctly must understand what a node and an antinode are and divide the wavelength by four. This understanding of what a node and an antinode are is also required to use a diagrammatic representation of a standing wave to determine the distance between the two. This is why the mistakes are penalized equally (-2 points).

    *d. Research questions*

We investigated the following research questions in the context of problems students solved:

RQ1: How do introductory students compare to graduate students (Ph.D. students in their first year) in their problem-solving performance?
RQ2: How do the different interventions impact students' problem-solving performance?
RQ3: How does drawing a diagram impact students' problem-solving performance?
RQ4: How facile are students at using the mathematical approach to solve the problem? What are some common difficulties students exhibit when using the mathematical approach?
RQ5: What are some common difficulties with students' use of diagrams to solve these problems?

Regarding productive diagrams, we recognize that how much value one derives from drawing a particular type of diagram and how the person employs the diagram (and the process of drawing it) to solve a problem depends on the expertise of the individual. However, for the purposes of this research, a diagram was considered productive if it could have aided students in solving the problem based upon a cognitive task analysis of the problem. The productive diagrams were classified in two broad categories: diagrams of third harmonics (whether correct or not) and diagrams of one wavelength (whether drawn as standing or single sinusoidal waves). A diagram from a student attempting to draw a third harmonic was considered to be productive even if it did not represent a third harmonic, or the third harmonic of the correct situation (tube open at one end and closed at the other). This is because these diagrams can be used to solve the problem by use of the expert-like approach. The second type of diagram (a diagram of one wavelength of a single sinusoidal/standing wave) was considered to be productive because it could be used to determine what fraction of a wavelength is the distance between a node and the adjacent antinode (the other type of productive diagram could be used to this end as well).

In order to answer the first four research questions and the first part of RQ4, we used the quantitative data collected from the introductory and graduate students and conducted an analysis of variance (ANOVA) [63]. To answer the second part of RQ4 and RQ5, we conducted interviews with eight students using a think-aloud protocol in order to obtain an in-depth account of their difficulties while solving the quiz problem and in addition provide some insights that would account for the performance of these students.

These students were at the time enrolled at the same university in an equivalent second semester algebra-based introductory physics course in which these concepts related to waves had been covered in the lectures and homework. There was no specific selection criterion other than the students being enrolled at the time in a second semester course, and the students varied in their course performance (in the first-semester course, their grades varied from a C to an A, and in the second semester course, in the latest exam prior to the interview, their grades varied from 56% to 83%) They had also been tested (via a midterm) on concepts related to waves before the interviews were conducted. We found that some of their homework assignments involved very similar problems to the ones analyzed in this study and their first midterm exam contained a problem requiring students to draw different harmonics of standing waves in tubes. The interviews were all conducted by the first author and during the interviews, students were first asked to solve the quiz problem to the best of their ability without interruption except they were asked to talk when they became quiet for a long time. After students were finished with the problem to the best of their ability, they were asked clarification questions if their reasoning at one point or another was unclear or questions related to other specific aspects of their problem-solving approach. Also, they were asked to solve the tube problem using another approach (i.e. if a student solved it using the mathematical approach he/she was asked if he/she can solve it using a diagrammatic approach and vice-versa). If students found it difficult to solve the problem using either approach, they would often be asked questions intended to provide scaffolding and guide them (some examples are provided in the results section). Each interview took between 1 and 1.5 hours (the students solved several other problems related to other studies) and were audiotaped. During each interview, the interviewer paid close attention to the student's solution approach and took careful notes throughout which kept track of what the student was writing and at what times. Whenever he noticed something worthwhile paying close attention to, he made a note of it along with the timestamp. This allowed the interviewer to revisit that particular part of the interview later having access to both the student's work (which was collected after the interview) and the audio recording in order to get a better understanding of what the student was doing and what particular difficulties he/she may have encountered. Those particular episodes were then written out with explanations of students' difficulties, and later both authors discussed them and decided which episodes were illustrative of the more common difficulties included in the article.

### III. Results

*RQ1: How do introductory students compare to graduate students (Ph.D. students in their first year) in their problem-solving performance?*

The quiz problem was also given to a group of 26 first year graduate students (physics experts for this study) enrolled in a TA training course. It is a straightforward mathematical exercise for a physics graduate student to solve for the wavelength using the mathematical approach. However, we found that 76% of them elected to draw a diagram to solve the problem (and ignored the equations provided to them completely), thus confirming our hypothesis that experts are more likely to follow the diagrammatic approach to solve this problem. The performance of both introductory physics students and graduate students on the quiz problem is listed in Fig. 2. The numbers on the bars represent the number of students in each group, and the error bars represent standard errors.

Pair-wise *t*-tests [63] comparing graduate students with introductory students reveal that graduate students who used the diagrammatic approach performed better than the introductory students who used the same approach ($p < 0.001$, Cohen's $d = 1.07$). Also, the overall scores of graduate students were better than the scores of introductory students ($p < 0.001$, Cohen's $d = 0.574$). For the comparison between the graduate students and introductory students who used the mathematical approach, a *t*-test is not appropriate due to the small number of graduate students in this group (10), but it does seem that there is little difference between the two groups.

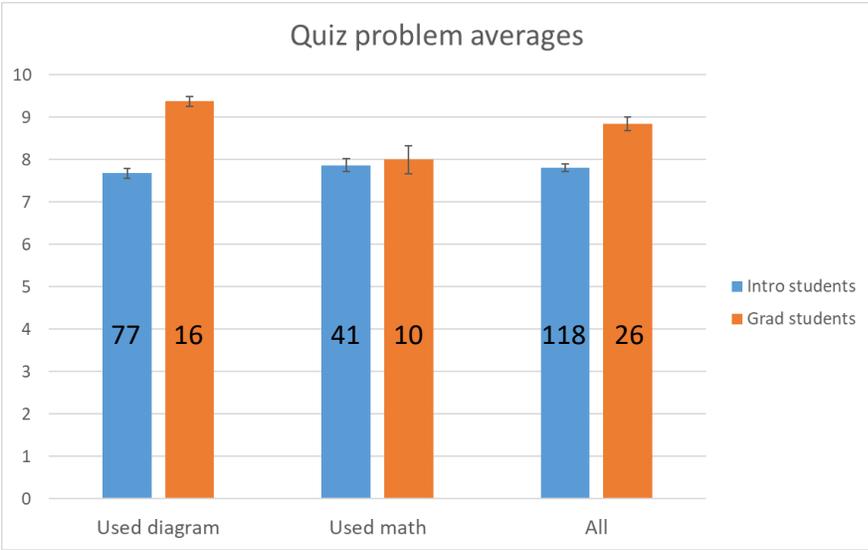

**Figure 2:** Performance of introductory and graduate students on the quiz problem by solution approach out of 10 points.

*RQ2: How do the different interventions impact students' problem-solving performance?*

The average scores on the quiz problem along with group sizes and standard deviations for the three different groups are shown in Fig. 3 (the numbers and error bars have the same meaning as in Fig. 2). ANOVA indicates that the groups show statistically significantly different performance and pair-wise t-tests between the different groups (corrected for multiple comparisons using the Scheffe method), shown in Table 3, indicate that students in DO performed significantly worse than students in the other two groups and that there is no statistically significant difference between the PO and NS groups.

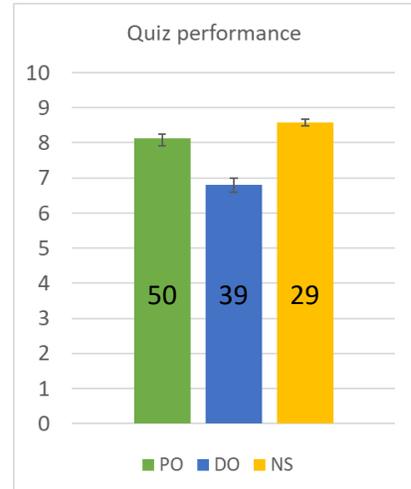

**Figure 3:** Average quiz performance of students in the different groups out of 10 points.

**TABLE 3:** p values for comparisons between the scores of the different groups controlling for multiple comparisons using the Scheffe method.

| Quiz | PO-DO | DO-NS | PO-NS |
|---|---|---|---|
| | 0.016 | < 0.001 | 0.414 |

*RQ3: How does drawing a diagram impact students' problem-solving performance?*

We investigated how drawing a diagram and/or using the diagrammatic approach vs. the mathematical approach impacted students' scores both in the quiz problem and in the midterm problem. All the students were placed in groups based on whether they used the more expert-like diagrammatic approach ("Used diagram" in Fig. 4) or primarily used the mathematical approach. Among the students primarily using the mathematical approach (as discussed earlier), we classified students in two categories based upon whether they drew a productive diagram or not. We wanted to investigate whether a productive diagram helped improve scores or not (hence, the students who used the mathematical approach were divided into two groups in Fig. 4: "Used math with productive diagram", and "Used math, no diagram").

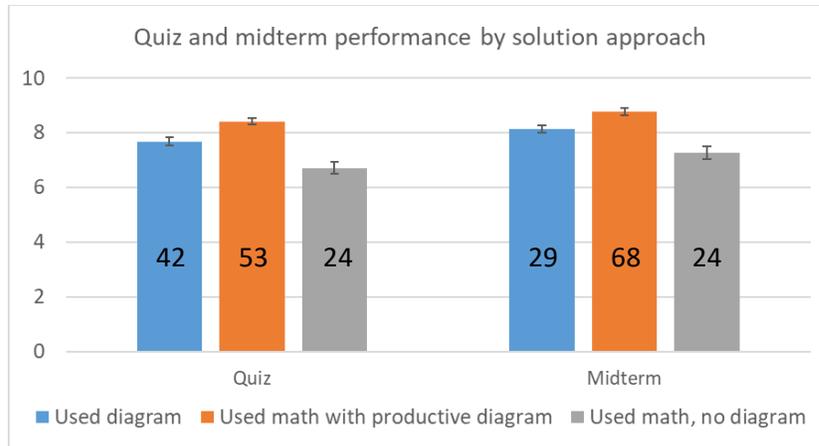

**Figure 4:** Average quiz and midterm performance of students separated by solution approach.

Comparison of the performance of students in the quiz with that on the midterm yields no statistically significant differences between any of the groups of students shown in Fig. 4 (students who used a diagram, students who used math, but also drew a productive diagram, students who used math without a productive diagram). However, both in the quiz and the midterm, students who primarily employed the mathematical approach but also drew a productive diagram performed better than students who chose the mathematical approach without drawing a productive diagram: p values/Cohen's d for comparing these two groups of students are, 0.002/0.900 and 0.006/0.697 in the quiz and the midterm, respectively. These values suggest that the difference between the performance of these two groups of students is quite large in both the quiz and the midterm.

*RQ4: How facile are students at using the mathematical approach to solve the problem? What are some common difficulties students exhibit when using the mathematical approach?*

Another finding is that introductory physics students who primarily used the mathematical approach had great difficulties in solving for the wavelength without plugging in a value for the speed of the wave, $v$. They were provided with the equation for the frequency of the $n^{\text{th}}$ harmonic of a wave in a tube open at one end, namely $f(n) = \frac{nv}{4L}$, but they were not given the relationship between the speed, frequency and wavelength of a wave ($v = \lambda f$). Therefore, students had to remember the equation $v = \lambda f$ in order to solve for the wavelength. Table 4 lists how many students, among those who wrote down $v = \lambda f$, were able to solve for the wavelength correctly without plugging in a value for the wave speed, how many students were not able to do so, and how many students plugged in some numerical value for the wave speed (despite the fact that it was not explicitly given) in order to solve for the wavelength both in the quiz problem and in the midterm problem.

    Table 4 shows that both in the quiz and the midterm, less than half of the students (48% in the quiz and 36% in the midterm exam) were able to eliminate the undesired quantities from the two equations and solve for the target variable without resorting to plugging in numerical information about wave speed that was not explicitly given. We note that the approach chosen by students in category 3 from Table 4 (plugging in a numerical value for the speed, $v$) is not necessarily an unproductive approach because it can help students reduce their cognitive load. However, the fact that so many students substituted a number for the wave speed in order to solve the quiz and midterm problems (when the wave speed would have canceled out between the two equations when solving for the wavelength) implies that a large fraction of the students in

the algebra-based introductory physics course are uncomfortable manipulating two equations symbolically in order to eliminate the undesired quantities and determine the target variable.

**TABLE 4:** Percentages of students who were able to solve for the wavelength algebraically, who were not able to do so, and who plugged in a value for the wave speed (although not provided) in order to solve for the wavelength (among the students who wrote down $v = \lambda f$) both in the quiz and in the midterm.

| **Quiz** | |
|---|---|
| 1. Solved correctly for $\lambda$ (algebraically, i.e. without plugging in a value for $v$) | 48% |
| 2. Did not solve correctly for $\lambda$ or did not solve at all | 10% |
| 3. Solved for $\lambda$ by plugging in a numerical value for $v$ | 41% |
| **Midterm** | |
| 1. Solved correctly for $\lambda$ (algebraically, i.e. without plugging in a $v$) | 36% |
| 2. Did not solve correctly for $\lambda$ or did not solve at all | 8% |
| 3. Solved for $\lambda$ by plugging in a numerical value for $v$ | 55% |

As noted by others, students are not always facile in transferring mathematical knowledge to a physics context. We examined whether students in algebra-based classes could solve an isomorphic (similar), purely mathematical problem. The problem is as follows:

In the two equations below, $C$ is a constant. Solve for $x$ in terms of $C$. Show your work!

$$\begin{cases} y = \dfrac{C \cdot z}{4} \\ z = x \cdot y \end{cases}$$

This is equivalent to the system of equations that students employing a purely mathematical approach must solve in the quiz and midterm problems if the following correspondences are made: $y \leftrightarrow f$, $C \leftrightarrow n$, $z \leftrightarrow v$, $x \leftrightarrow \lambda$, and $4 \leftrightarrow 4L$ ($L$ was given as 1.5 m so students could plug it in to get $4L = 6$). We find that 64% of students at the beginning of the first semester algebra-based course are able to solve this system correctly and 89% of students at the beginning of their second semester course are able to solve this system correctly. It may not be appropriate to compare these percentages with the percentages of students who solved the quiz and midterm exam problems using the mathematical approach algebraically, without plugging in a value for $v$, because in those cases, students had the option of plugging in a value for one of the unknowns ($v$) which greatly simplifies the task. It seems reasonable to expect that some of the students (if not the majority) who plugged in a value for $v$ in the equation did so because otherwise, they would have been unable to solve the problem. However, it does appear that students are more adept at solving for the desired variable from the system of two equations in the purely mathematical context than in the physics context, especially students in the second semester algebra-based class.

These difficulties were also observed in the interviews. Some students approached the problem mathematically at first, but then changed their approach to diagrammatic when they had difficulty determining what mathematical steps to perform next. Dan, for example plugged in $n = 3$ and $L = 1.5$ m in the equation for frequency and solved for $f/v$ to get 1/2 (he did not write down the units of 1/m). At this point, he appeared to be stuck and after some thinking, he changed his approach to the diagrammatic one. After the think-aloud part was over and Dan was probed further, he noted that he was aware of the other equation, $v = f\lambda$. He noted that at one point while solving the problem he thought about using this equation to find the wavelength. However, he did not explicitly write it down because he was not sure if it would help him to determine the wavelength. In particular, after he switched to solving the problem using the diagrammatic approach and attempted to solve it to the best of his abilities, the interviewer asked him if he was aware of the connection between speed of a wave, frequency and wavelength. At this point, Dan wrote this equation on the paper and identified correctly that the wavelength would equal 2 m. It was interesting that Dan noted that in his mind he had tried to think if he could solve for the wavelength using this equation $v = f\lambda$ along with $f/v = 1/2$ when he was solving the problem during the think aloud part of the interview without probing. However, he gave up on trying to solve the problem using these equations and did not realize that writing down the equation $v = f\lambda$ on paper may have reduced the cognitive load during problem solving and may have helped facilitate the problem-solving process. Furthermore, while Dan himself noted that he contemplated using the equation $v = f\lambda$ along with $f/v = 1/2$ to solve for the wavelength; he was unable to do this and resorted to another approach. However, when he was given a system of two equations (without a physics context) with two unknowns of the form $\begin{cases} x + 2y = 3 \\ 3x - y = 2 \end{cases}$, he was able to readily solve this purely mathematical problem for variables *x* and *y* without much effort.

Another student, Karen, initially solved the problem using the diagrammatic approach during the think aloud part of the interview. During the second part of the interview, the interviewer asked her to solve the problem using the mathematical approach and gave her the equation $v = f\lambda$. At this point, Karen's first step was to substitute this equation into the other equation provided with the problem and plug in $n = 3$ and $L = 1.5$ m. She thus obtained $f = \frac{3(f\lambda)}{4(1.5)}$ which was a productive way to solve the problem. However, after this step, she was unsure about what to do and after some thinking, she gave up and noted that she did not know how to proceed (she did not realize that the frequency can be canceled from both sides of the equation). The interviewer then gave her a system of two equations with two unknowns (traditional *x* and *y* variables without the physics context) similar to the one Dan had to solve, and she was able to solve it correctly without much effort. In this situation, Karen was aware of what needed to be done next, but in the tube problem situation, after the substitution step, she was not able to determine what to do next. She did not realize that *f* was on both sides of the equation and she could cancel it or that she could multiply both sides of the equation by the denominator of the fraction on the right to get a simpler equation for the wavelength.

Another student, Tara, approached the problem mathematically from the beginning. She knew the two equations that needed to be used, $f = \frac{nv}{4L}$ and $v = f\lambda$, wrote them down, and then said the following:

Tara: "If I knew *v* [speed of the wave] I could plug in this equation $[f = \frac{nv}{4L}]$, get *f* and then plug that in this equation [$v=f\lambda$] to get the wavelength."

She then drew one wavelength of a travelling wave and said that she would divide the wavelength that she obtains by 4 to get the distance between the node and antinode. At this point she indicated that she was done to the best of her ability and her statement indicated that she could not solve this problem since the speed of the wave was not given. At this point, the interviewer then asked her:

Interviewer: "Could you do this without knowing what *v* is?"

Tara: "Could I? Probably..."

She then thought about how she could solve for the wavelength using the two equations for some time (a little less than a minute) and said:

Tara: "I can't think of another way."

Interviewer: "If you look at this equation [pointing with finger to $f = \frac{nv}{4L}$] and this equation [pointing with finger to $v = f\lambda$]…"

Tara interrupted the interviewer before he could ask the question "could you solve for $\lambda$?":

Tara: "I can plug it all in […] use substitution […] you would plug the frequency and the wavelength in for *v* [she meant plug in the frequency *times* the wavelength for *v*] in the equation given […] so you can solve for lambda that way."

She then correctly solved for the wavelength without plugging in a value for the speed of the wave. It appears that even though Tara knew how to perform substitution algebraically, she did not retrieve this information even after being explicitly asked if she could solve the problem without plugging in a value for the speed. Moreover, the fact that after the interviewer directed her attention to the two equations that had to be manipulated, Tara realized immediately what she needed to do, suggests that when she earlier paused (for about a minute) to think about whether she could solve the problem without knowing the speed, she may have not been focusing on the relevant information about the two equations.

These examples from interviews suggest that students were able to solve two simultaneous equations without a physics context without any difficulty but there was a lack of transfer of the mathematical knowledge to a physics context. This difficulty in transferring learning from the mathematical context to the physics context could be, e.g., due to the fact that in the physics context there may be other information which can distract students from processing the relevant information, while such distractions are not present when engaged in a purely mathematical exercise. Moreover, mathematics is used differently in physics courses from mathematics courses and this could also lead to difficulties in transfer from one context to another. For example, solving for the wavelength from the two equations, $f = \frac{nv}{4L}$, and *v* = *f*λ may be more difficult for a student than solving a general system of two simultaneous equations with *x* and *y* variables (as observed with some students in the interviews). In the first case, the symbols that go into those equations have physical meanings and it may be more difficult for students to focus on the relevant information (e.g., substitute one equation in the other, cross-multiply etc.) because the physical meanings add more information that is not present in the second case in which the variables *x* and *y* are devoid of physical meaning (the same can be said about the isomorphic mathematical problem which included variables *x*, *y* and *z*). This could partly account for the difficulties students exhibited in solving for the wavelength from the two equations without resorting to plugging in information that is not given, observed in the quantitative data and in the think-aloud interviews.

*RQ5: What are some common difficulties with students' use of diagrams to solve these problems?*

Interviews also revealed some difficulties algebra-based introductory physics students encountered while using the diagrammatic approach to standing waves in the tube while solving the problem. Karen, for example, stated at the beginning after reading the problem that at the closed end of the tube there will be a node and that at the open end of the tube, there will be an antinode. She then tried to draw the third harmonic of this wave, and her attempts reflected this knowledge. However, she had difficulty drawing the third harmonic directly and she decided to start with a drawing of the first harmonic on the side (not in the tube) and work up to the third. However, the diagrams of the harmonics she drew on the side had nodes at both ends and therefore corresponded to a different situation (tube closed at both ends). Karen was unaware of this mistake in her drawing despite the fact that she explicitly stated at the beginning after reading the problem statement that at one end of the tube there should be a node and at the other end there should be an antinode. When solving the problem, she appeared to have forgotten about her initial correct statement (there should be a node at one end and antinode at the other end) and used the incorrect third harmonic she drew on the side (which corresponded to a standing wave in a tube closed at both ends) to solve the problem.

Another student, Sara, drew the 5$^{th}$ harmonic for the wave in the tube open at only one end. However, the last section of the wave she drew (last 1/4 wavelength) looked on her diagram (reproduced in Fig. 5) to be of the same length as the other two sections of the wave (1/2 wavelength).

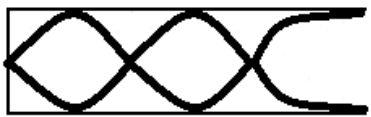

**Figure 5:** Diagram of the fifth harmonic as drawn by a student, Sara (reproduction).

Consequently, she divided the length of the tube by three to get the distance between two nodes and then divided that distance by two to get the distance between a node and an antinode. After she was satisfied with her answer and was done with the problem to the best of her ability, the interviewer asked her why she divided the length of the tube by three. Here is a short excerpt:

Sara: "Well, from what it said about third harmonic, I drew waves so you get one node here, another node here [the two middle nodes] and then the rest of the wave just opens up to the outside of the tube. Assuming the tube was closed, I think you would get another node right at the end of the tube [right side]."

Interviewer: "Yeah, but it's not closed."

Sara: "Right, but I assumed that these nodes [the two middle ones] would automatically split the tube into three."

Interviewer: "Okay, so you're thinking that here as well [at the right end] you would have a node?"

Sara: "If it was closed […] I know it's not closed so you don't get the node, it just kind of opens out, but I assumed, if it was closed you would get that node and these nodes [the two middle ones] would split the tube into three equal […] lengths."

Instead of correcting her diagram to fit the problem situation, Sara seemed to have modified the problem to fit her diagram and essentially ended up solving a different problem. She was also aware that it was a different problem ("I know it's not closed") but this didn't seem contradictory enough to her to change her diagram or interpretation. At this point the interviewer continued with further questioning to draw attention to her mistake:

Interviewer: "Sure, but you're solving a different problem, because you're assuming it's closed and it's not. What would change if it's open?"

Sara: "So basically this [the last section on the right which should have been ¼ wavelength] is not the same as this and this [she pointed to the other two sections of the wave she drew]".

Sara then stopped to think about her diagram and used the knowledge that the last section is half the length of the other two sections to correctly solve for the distance between a node and the adjacent antinode. Similar to Karen, Sara also did not use knowledge she possessed (the last section of the wave was shorter than the other two sections) when she initially solved the problem without interruption. After the interviewer explicitly pointed her attention to the diagram she drew and pointed out that she had solved a different problem, Sara was able to retrieve the correct information and use it to solve the problem correctly based on her diagram of the fifth harmonic.

Another student, Brian, used information that was not applicable in the quiz problem. He thought that the distance between a node and an antinode decreases as you move away from the closed end. The diagram Brian drew is shown in Fig. 6 and has a standing wave in which the distance between nodes decreases away from the closed end of the tube. It is unclear why he thought this to be true, but he explicitly stated that he remembered his instructor drawing a diagram where this was the case. It is unlikely that the instructor drew such a diagram because the book the students used had no such diagram or any discussion of a situation where the distance between nodes of a standing wave changes. It is possible that he misinterpreted a diagram drawn by the instructor.

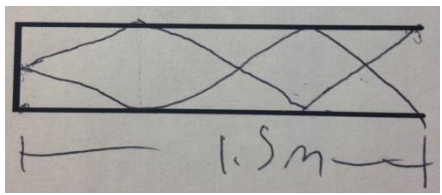

**Figure 6:** Diagram of the fifth harmonic drawn by Brian (a student).

Brian did not realize that the diagram he drew cannot be correct for the situation presented in the problem because if his reasoning was correct, then the problem should have specified *which* node to adjacent antinode distance students had to find (this distance would be different for his situation depending on which node/antinode you choose). He was therefore unable to solve the problem using this diagram and proceeded to try the mathematical approach.

*Other findings from interviews*

One of the main findings from the quantitative investigation is that a good diagram is valuable for solving a problem related to a standing wave in a tube even when a student employs a primarily mathematical approach to problem solving. In particular, we found that students who used a mathematical approach but drew a productive diagram performed better than students who used the mathematical approach without drawing a productive diagram. As noted earlier, when defining "productive diagram" for these problems, it was considered that any diagram of a third harmonic (whether or not correct) could be productive because it gives students an opportunity to perform a conceptual analysis and planning related to the problem and students could use the insight derived from drawing this diagram to solve the problem. Moreover, even in the case when primarily a mathematical approach was chosen, the process of drawing a productive diagram can be helpful in conceptually analyzing the problem and such a diagram could be used to determine what fraction of a wavelength was represented by the distance between a node and the adjacent antinode. Another type of diagram that could be useful and was considered productive was a diagram of one wavelength of a standing or single sinusoidal wave. Interestingly, half of the students interviewed (four) chose the mathematical approach. Most of these students (three) drew a diagram of one wavelength of a single sinusoidal wave in order to determine that the distance between a node and the adjacent antinode is one quarter of the wavelength (and did so correctly). One of the interviewed students, drew a diagram of the third harmonic but did not explicitly use the diagram she drew in solving the problem (and only focused on the equations). This student divided the wavelength by three (instead of four) to obtain the distance between the node and the antinode because she claimed that the number she needed to divide the wavelength by in order to determine the distance between a node and the adjacent antinode was related to the harmonic (i.e., she divided by *three*, because the problem involved the *third* harmonic of a standing wave).

Among the four students who used the diagrammatic approach, only one used it in the way that an expert would most likely use it (and consistent with the approach of graduate students). He determined how many distances between a node and an antinode would fit in the length of the tube and then divided the length of the tube by that number. The other three students used the diagram of the third harmonic they had drawn to determine the wavelength. After finding the wavelength, they were at the same point as the students who used the mathematical approach to determine the wavelength, and just like those students, they then proceeded to determine the number by which to divide the wavelength in order to get the distance between a node and the adjacent antinode. To this end, one of these three students drew an additional diagram of one wavelength of a single sinusoidal wave (and used it incorrectly to obtain the distance between a node and an antinode) while the other two students explicitly used the diagrams of the third harmonic they had drawn (and used the diagrams correctly to find the distance between a node and an antinode). What the interviewed students did on their own while solving the problem and thinking aloud and what they said when asked for clarification of the points they had not made earlier suggests that drawing a diagram of a third harmonic for the problem or even one wavelength of a single sinusoidal wave can be helpful in finding the relationship between the distance between a node and the adjacent antinode and the wavelength of a standing wave because it helps students focus on relevant information in order to proceed with the problem solution. As noted earlier, the interviewed student who neither used her diagram of the third harmonic nor drew one wavelength of a single sinusoidal wave to determine the distance between the node and antinode with respect to the wavelength made a mistake (divided the wavelength by three because the problem involved the third harmonic). However, out of the other students who either used their diagrams of the third harmonic or diagrams of one wavelength of a single sinusoidal wave to determine the distance between the node and antinode, only one made a mistake (divided the wavelength by 2 instead of 4). The interviews suggest that students who drew productive diagrams performed better even if their chosen approach was primarily mathematical because the diagram helped them think about the problem solution conceptually.

### IV. Summary and Discussion

In the first part of this two-part article, we reviewed the research on problem-solving in physics which can be broken down into three inter-related categories: information processing and cognitive load, knowledge organization, and metacognition and problem-solving heuristics. In this part 2 of the two-part article, we discussed a research study which speaks directly to several important aspects of problem-solving. We found that among the students who chose primarily the mathematical approach, those who drew productive diagrams performed better than those who did not (RQ3). This result is consistent with prior research which found that students who draw diagrams outperform students who do not even in a multiple-choice exam [35]. The study presented here was not designed to clarify whether the students who drew diagrams were the ones who generally had more expert-like approaches to problem solving (e.g., an approach which includes a conceptual analysis stage that started with or involved drawing a diagram) or whether the process of drawing a diagram helped students regardless of their general problem-solving approaches. However, it is important to note that the interviews suggested that students who did draw diagrams were attempting to make sense of the problem conceptually and that the students who explicitly used the diagrams they drew were less likely to make mistakes than the students who did not draw a diagram. Thus, encouraging students and helping them explicitly learn good problem-solving heuristics which include drawing a diagram in the conceptual planning stage would benefit them, and the instructors should emphasize and reward students for drawing diagrams.

We note that the quiz problem was also administered to a set of graduate students for two reasons: to confirm that the more expert-like approach is indeed the diagrammatic approach and also to obtain a benchmark for what would be the upper-limit of the performance of introductory physics students. We found (RQ1) that the majority of graduate students selected the diagrammatic approach even when the equation for the $n^{th}$ harmonic frequency was provided, thus confirming that the diagrammatic approach is indeed a more expert-like approach. This finding supports prior research which has emphasized that experts tend to make significant use of representations when engaged in problem solving [1-6, 9, 13, 16, 32, 33]. We also found that graduate students outperformed introductory physics students by an average of about 13%.

Furthermore, we found that the students who were given a diagram of an empty tube performed significantly worse than both the students who were asked to draw a diagram and the students who were not given any instructions regarding diagrams (RQ2). In a previous investigation [64], we found the same result while examining introductory students' performance on two problems in electrostatics that involved considerations of initial and final situations. In addition, for the electrostatics problems, the differences in score between group DO and the other two intervention groups were even more pronounced (average of students given diagrams was more than 20% lower than the averages of the other two groups and the p values were less than 0.001). The research in Ref. [64] involved the same methodology as that described here. However, the diagrams given to students in group DO were very similar to what most instructors would initially draw in order to solve those two problems and were intended as scaffolding support. Instead of helping students solve those two problems, the given diagrams had the opposite effect, statistically worsening their performance as compared to students in the other two groups. Heckler [65] conducted a similar study and found that students who were provided with diagrams tended to follow more formal methods as opposed to use their intuition (which was sometimes productive for some of the problems), and this caused deteriorated performance compared to a control group for some of the problems. Although in other cases, no differences in performance were observed. Unlike the diagrams provided in Refs. [64] and [65], in the study presented here, students in group DO were given only a partial diagram (empty tube). Providing students the partial diagram was intended as a hint or prompt for them to complete it and attempt to solve the problem in an expert-like manner (drawing a diagram of a third harmonic of a standing wave and using it to solve the problem). However, similar to the study involving electrostatics problems, here we also find that providing the diagram of the empty tube had the opposite effect from what was intended. Other researchers have found [66] that providing diagrams to students provides a marginal benefit, and that in cases where diagrams are not provided, students are often significantly more likely to draw their own diagrams while engaged in problem-solving, suggesting that it may be beneficial for instructors to avoid providing diagrams for their students (and instead ask students to draw them explicitly, or reward them for doing so), except in cases where the problem situation may be complex and difficult to explain using only words.

We also found that in the context of solving the physics problem, students had great difficulty manipulating two equations symbolically (RQ4). However, when students had to solve an isomorphic mathematical system of two equations devoid of physics context, the vast majority of them were able to manipulate and solve for the target variable correctly in terms of other variables. This discrepancy between students' mathematical ability in a physics context and their mathematical ability in a mathematical context was also observed in the interviews, and is consistent with prior research on student difficulties when using mathematics in a physics context [67], which often points to the fact that the way in which math is used in physics courses is very different from how it is used in mathematics courses [68, 69]. Many students expressed the need to substitute a numerical value for the speed of the wave before they solved for the wavelength using the two simultaneous equations even though the speed would have canceled out between the two equations. When they were asked to solve a system of two simultaneous equations in a purely mathematical context, all the students were able to do so with little effort, thus suggesting that the physical context in which the mathematical task was encased in the problem given in the quiz and midterm may have made it more challenging for students to recognize how to manipulate the equations to solve for the target variable. It is possible that students may not be focusing on the relevant information due to the additional context (the variables in the physics problem have meaning which adds additional information that must be ignored when performing the necessary mathematical manipulations). Interviews suggested that in the initial problem-solving phase, some students did not retrieve mathematical knowledge relevant to algebraically manipulate two simultaneous equations, even though they knew how to do it. It is possible that as a result of their expertise level, the physics context had too much information to be processed at a given time in their working memory and caused cognitive overload [70]. Consequently, it was more difficult for them to focus on the relevant information that had to be processed at one time and make productive decisions in order to move forward with the solution. For example, as discussed earlier, one interviewed student was unable to determine how to solve the problem without plugging in information about the speed of the wave, which was not given, even after the interviewer explicitly asked her to do so. However, once the interviewer directed the student's attention to the two equations that had to be manipulated, that student immediately realized the next step (algebraic substitution) and solved the problem correctly.

In addition, in the interviews, difficulties were also observed when students were engaged in solving the problem using the diagrammatic approach (RQ5). Some of these difficulties could also be interpreted using the framework of cognitive load theory [70]. In particular, sometimes students did not make use of knowledge about waves that they possessed, which was explicitly mentioned by the student and, at times, even used briefly by the same student at another stage of problem solving. Interviews suggest that at various points in problem solving, students had cognitive overload while they focused on certain aspects of the problem, and they completely lost track of other important information which led to deteriorated performance.

One instructional implication of this research is that students should be encouraged to draw productive diagrams by rewarding them for drawing them. One of the many frameworks that may be useful for helping students learn to draw productive diagrams and other effective approaches to solving physics problems is the field-tested cognitive apprenticeship model. Within this cognitive apprenticeship model, the instructor can model productive diagrams while exemplifying effective approaches to problem solving, then coach students and provide feedback while they practice these skills and then gradually remove the support as they develop self-reliance. Another instructional implication of this investigation is that it is important for instructors to keep in mind that algebra-based introductory physics students can have cognitive overload while solving physics problems as they must manage both the mathematical manipulations and how to use the underlying physical principles simultaneously to proceed successfully in the vast problem space. Trying to juggle both tasks at the same time can be cognitively demanding particularly for introductory physics students in algebra-based courses who are not facile in algebra. A major fraction of their working memory may be used either in comprehending the mathematical procedure or in processing the related physics concepts. For example, students whose significant cognitive resources are allocated to parsing the mathematics involved rather than in sense making of the underlying physics principles and *why* certain concepts were used, may find it difficult to build a good mental picture of the concepts involved and may have difficulty in solving physics problems successfully. Since mathematical difficulties can make it challenging for students to build a good knowledge structure of physics, appropriate guidance and suitable scaffolding support should be provided to students, accounting for their physics and mathematics competencies in order to take students gradually from their initial knowledge state to the final knowledge state based upon the goals of the course.

Given two of the findings in this study which strengthen prior research as noted earlier (namely that 1. students who draw diagrams perform significantly better than students who do not even if their approach is primarily mathematical and that 2. providing students with a diagram can have a detrimental effect on their problem-solving performance), future research could strengthen these results by investigating the impact of prompting students to draw diagrams and rewarding them over the course of a semester on their problem-solving skills. In this study, the instructions that students received regarding diagrams varied and thus no cumulative effects of students receiving the same instructions regarding diagrams could be observed, and it would beneficial to learn what those cumulative effects could be if the students remained in one of the three groups throughout the semester. Also, students were not rewarded for drawing diagrams or using effective problem-solving strategies, and since assessment drives learning [71], rewarding students via grade incentives for drawing diagrams could have positive effects on their problem-solving performance. Another of our main findings from the interviews is the usefulness of the idea of cognitive load in understanding students' mathematical difficulties. As of now, there isn't a lot of research in physics education investigating students' mathematical difficulties in a physics context using a cognitive lens. Given the emphasis that prior research has put on investigating problem-solving in general from the perspective of information processing and cognitive load, we believe that there is a lot to be gained in terms of understanding students' mathematical difficulties (in the context of problems-solving in physics) from research involving this perspective. Future research investigating this could provide important insights that may have instructional implications for educators interested in helping students develop effective problem-solving skills in physics.